\documentclass[epj,referee,onecolumn]{svjour}
\usepackage{latexsym}
\usepackage{amsmath}
\usepackage{amssymb}
\usepackage[dvips]{graphicx}
\usepackage{color}
\begin{document}
\title{MHD waves and instabilities in flowing solar flux-tube plasmas in the framework of Hall magnetohydrodynamics}
\author{I.~Zhelyazkov 
\thanks{\emph{e-mail:} izh@phys.uni-sofia.bg}%
}                     
%
%
\institute{Faculty of Physics, Sofia University, BG-1164 Sofia, Bulgaria}
\date{Received: date / Revised version: date}
%
\abstract{
It is well established now that the solar atmosphere, from photosphere to the corona and the solar wind is a highly structured medium.  Satellite observations have confirmed the presence of steady flows.  Here, we investigate the parallel propagation of magnetohydrodynamic (MHD) surface waves travelling along an ideal incompressible flowing plasma slab surrounded by flowing plasma environment in the framework of the Hall magnetohydrodynamics. The propagation properties of the waves are studied in a reference frame moving with the mass flow outside the slab.  In general, flows change the waves' phase velocities compared to their magnitudes in a static MHD plasma slab and the Hall effect limits the range of waves' propagation.  On the other hand, when the relative Alfv\'enic Mach number is negative, the flow extends the waves propagation range beyond that limit (owing to the Hall effect) and can cause the triggering of the Kelvin--Helmholtz instability whose onset begins at specific critical wave numbers.  It turns out that the interval of Alfv\'enic Mach numbers for which the surface modes are unstable critically depends on the ratio between mass densities outside and inside the flux tube.
\PACS{
      {96.50.Tf}{MHD waves; plasma waves, turbulence }   \and
      {96.50.Ci}{Solar wind plasma; sources of solar wind}
     } 
} 
\titlerunning{Hall-MHD waves and instabilities in flowing solar plasmas}
\maketitle
\clearpage
\onecolumn
\section{Introduction}
\label{sec:intro}
Various waves and oscillations which occur in structured solar atmosphere were intensively studied over the past three decades and an exclusive review of their theory the reader can find in \cite{valery07} and references therein.  Next step in studying the wave phenomena in solar and stellar atmospheres was the consideration of steady flows there.  Satellite measurements of plasma characteristics of, for instance, the solar wind and coronal plumes flows, such as the magnetic field, the thermal and flow velocity and density of plasma or plasma compositions, are important to understand the various plasma wave modes which may arise. However, wave analysis requires further information and special tools to identify which set of modes is contributing to observed wave features.  In practice, one may use filters to perform the so-called \emph{pattern recognition\/} to detect the various kind of modes that may propagate in plasma and to determine their contribution to the wave energy \cite{vocks99}.  Another important issue is the question for the waves' stability.  The magnetosonic waves in structured atmospheres with steady flows have been examined by Nakariakov and Roberts \cite{valery95}, Nakariakov et al.\ \cite{valery96}, Andries and Goossens \cite{andries01}, and Terra-Homem et al.\ \cite{terra03}.  Andries and Goossens studied also the conditions at which resonant flow and Kelvin--Helmholtz instabilities take place, and Terra-Homem et al.\ have investigated the effect of a steady flow on the linear and nonlinear wave propagation in various solar structures.

It is worth mentioning that all the aforementioned studies were performed in the framework of standard magnetohydrodynamics.  It was Lighthill \cite{lighthill60} who pointed out almost $50$ years ago that for an adequate description of wave phenomena in fusion and astrophysical plasmas one has to include the Hall term, $m_{\rm i} (\mathbf{j} \times \mathbf{B})/e\rho$, in the generalized Ohm's law.  That approach is termed Hall magnetohydrodynamics (Hall MHD).  In this way, it is possible to describe waves with frequencies up to $\omega \approx \omega_{\rm ci}$.  Because the model still neglects the electron mass, it is limited to frequencies well below the lower hybrid frequency: $\omega \ll \omega_{\rm lh}$.  Generally speaking, the theory of Hall MHD is relevant to plasma dynamics occurring on length scales shorter than an ion inertial length, $L < l_{\rm Hall} = c/\omega_{\rm pi}$ (where $c$ is the speed of light and $\omega_{\rm pi}$ is the ion plasma frequency), and time scales of the order or shorter than the ion cyclotron period ($t < {\omega_{\rm ci}}^{-1}$) \cite{huba95}.  Thus the Hall MHD should affect the dispersion properties of MHD waves in spatially bounded magnetized plasmas.

The first paper devoted to the propagation of fast MHD and ion-cyclotron surface waves at a plasma--vacuum interface in the limit of cold plasma was that by Cramer and Donnelly \cite{cramer83}.  Later on, Cramer \cite{cramer91} generalizes that model obtaining the dispersion of nonlinear surface Alfv\'en waves.  It is worth noticing that even in unbounded magnetoplasma at purely parallel propagation the Alfv\'en waves become dispersive when the Hall current is taken into account in the basic MHD equations \cite{almaguer92}.  In Ref.~\cite{ivan96} the Hall MHD has been applied in studying the parallel propagation of fast wave in an ideal static plasma slab.  There were derived four boundary conditions (see Sec.~III in \cite{ivan96}) necessary for obtaining the dispersion relations of sausage and kink modes in spatially bounded magnetized plasmas.  That approach has been applied in studying the oblique propagation in the same geometry both for low-$\beta$ and finite-$\beta$ plasmas \cite{ivan00,ivan03}.

The first study on surface-wave parallel propagation in a flowing ideal MHD flux tube surrounded by a static plasma environment (both embedded in a constant magnetic field $\mathbf{B}_0$) in the framework of the Hall MHD was performed by Miteva et al.~\cite{rossi03}.  It has been shown that while in a static plasma slab the hydromagnetic surface waves (sausage and kink modes) are Alfv\'en ones (their phase velocities are close to the Alfv\'en speed in the layer), in slabs with steady flows they become super-Alfv\'enic waves. Moreover, as it is logical to expect, there exist two type of waves: forward and backward propagating ones, bearing in mind that the flow velocity defines the positive (forward) direction.  An in-deep examination of wave modes in flowing solar-wind-flux-tube magnetized plasmas for finite-$\beta$ and zero-$\beta$ ionized media has been performed in \cite{rossi04,sikka04} (see also the references in those papers).

Hall MHD is relevant not only to linear MHD waves, but also to nonlinear ones \cite{michael02a,michael02b,ballai03,mahajan05}.  Dispersive effects caused by Hall currents perpendicular to the ambient magnetic field can influence the generation and propagation of shock waves \cite{ballai07}.  Recently Clack and Ballai \cite{clack08} studied the nonlinear wave propagation of resonant slow magnetoacoustic waves in plasmas with strongly anisotropic viscosity and thermal conductivity alongside of dispersive effects due to Hall currents.  They show that the nonlinear governing equation describing the dynamics of nonlinear resonant slow waves is supplemented by a term which describes nonlinear dispersion and is of the same order of magnitude as nonlinearity and dissipation.  In the case of stationary nonlinear waves the Hall-MHD equations can be rewritten in the so-called Sakai--Sonnerup set of equations that describe the plasma state and provide oscillatory and solitary types of solutions \cite{rossi08}. The overall parameter study on the polarization characteristics, together with the magnetic field components and density variations in different ranges of solutions performed in that paper might be further on applied to the theoretical treatment of particle interaction with such waves, e.g., at shocks in space plasmas, possibly leading to particle acceleration.  MHD parametric instabilities of parallel propagating incoherent Alfv\'en waves are influenced by the Hall effect and that is especially important for the left-hand polarized Alfv\'en waves \cite{nariyuki07}.  Ruderman and Caillol \cite{michael08} claim that the left-hand polarized Alfv\'en waves in a Hall plasma are actually subject to three different types of instabilities, namely modulational, decay, and beat instability, while the right-hand polarized waves are subject only to decay instability.  A new trend in the Hall MHD is its application to partially ionized plasmas \cite{pandey08} -- the Hall effect may play an important role in the dynamics of weakly ionized systems such as the Earth's ionosphere and protoplanetary discs.  The exact nonlinear cylindrical solution for incompressible Hall-MHD waves, including dissipation, essentially from electron--neutral collisions, obtained in a uniformly rotating, weakly ionized plasma, derived by Krishan and Varghese \cite{krishan08}, demonstrates the dispersive nature of the waves, introduced by the Hall effect, at large axial and radial wave numbers.  Such short-wavelength and arbitrarily large-amplitude waves could contribute toward the heating of the solar atmosphere.  One of the most important mechanisms for solar atmosphere's heating and especially the solar wind is the wave turbulence, which is also affected by the Hall currents \cite{krishan04,galtier06,galtier07,galtier09,shaikh09}.  The strongest competitor of the wave turbulence heating mechanism, notably the magnetic reconnection, is similarly heavily influenced by the Hall effect \cite{ma01,morales05,arber06,cassak07,craig08}.  One even speaks for Hall reconnection being akin to Petschek reconnection model.  It is worth mentioning that the Hall effect can influence the magnetic field dynamics in dense molecular clouds \cite{wardle99}, magnetorotational instability \cite{sano02}, star formation \cite{wardle04}, as well as compact objects \cite{elena08}.  As seen, the Hall MHD has impact on many important astrophysical phenomena and objects.

Here, we investigate the influence of flow velocities on the dispersion characteristics and stability of hydromagnetic surface waves (sausage and kink modes) travelling along an infinitely conducting, magnetized jet moving past also (with a different speed) infinitely conducting, magnetized plasmas.  If in the solar corona plasma $\beta$ (the ratio of gas to magnetic pressure) is much less than unity, in the solar wind flux tubes it is $\beta \approx 1$.  Since we are going to study the wave propagation in flowing solar wind plasma, we can assume that we have a `high-$\beta$' magnetized plasma and treat it as an incompressible fluid.  For simplicity, we consider a planar jet of width $2x_0$ (embedded together with environments in a constant magnetic field $\mathbf{B}_0$ directed along the $z$ axis), allowing for different plasma densities within and outside the jet, $\rho_{\rm o}$ and $\rho_{\rm e}$, respectively.  The most natural discontinuity which occurs at the surfaces bounding the layer is the tangential one because it is the discontinuity that ensures a static pressure balance \cite{cravens97}.  For typical values of the ambient constant magnetic field $B_0 = 5 \times 10^{-9}$ T and the electron number density inside the jet $n_{\rm o} = 2.43 \times 10^6$ m$^{-3}$ at 1 AU \cite{cravens97}, the ion cyclotron frequency $\omega_{\rm ci}/2\pi = 76$ mHz, and the Hall scale length ($= v_{\rm Ao}/\omega_{\rm ci}$, which is equivalent to $c/\omega_{\rm pi}$) is $l_{\rm Hall} \approx 150$ km.  This scale length is small, but not negligible compared to layer's width of a few hundred kilometers.  Here, we introduce a scale parameter $\varepsilon = l_{\rm Hall}/x_0$ called the \emph{Hall parameter}.  In the limit of $\varepsilon \to 0$, the Hall MHD system reduces to the conventional MHD system.  Our choice for that parameter is $\varepsilon = 0.4$.  The flow speeds of the jet and its environment are generally rather irregular.  For investigating the stability of the travelling MHD waves it is convenient to consider the wave propagation in a frame of reference attached to the flowing environment.  Thus we can define the relative flow velocity $\mathbf{U}^{\rm rel} = \mathbf{U}_{\rm o} - \mathbf{U}_{\rm e}$ ($U_{\rm o}$ and $U_{\rm e}$ being the steady flow speeds inside and outside the flux tube, respectively) as an entry parameter whose value determines the stability/instability status of the jet.  As usual, we normalize that relative flow velocity with respect to the Alfv\'en speed in the jet and call it Alfv\'enic Mach number $M_{\rm A}$, omitting for simplicity the superscript ``rel''.  Another important entry parameter of the problem is $\eta = \rho_{\rm e}/\rho_{\rm o}$.  It turns out that the waves' dispersion characteristics and their stability critically depend on the magnitude of $\eta$.  Our preliminary numerical studies, based on a criterion for rising the Kelvin--Helmholtz instability (namely that the modulus of the steady flow velocity must be larger than the sum of Alfv\'en speeds inside and outside the flux tube) proposed by Andries and Goossens \cite{andries01}, showed that in order to expect instability onset at some reasonable values of the relative Alfv\'enic Mach number $M_{\rm A}$, one should assume fairly large values of the parameter $\eta$.  In this study we take $\eta = 10$ which means that the Alfv\'en speed inside the jet is roughly three times larger than that in the environment -- actually an ``edge'' choice, but still acceptable.  Thus the waves' dispersion characteristics (the dependence of the wave phase velocity $v_{\rm ph} = \omega/k$ on the wave number $k$) and their stability states are determined by the three parameters $\eta$, $\varepsilon$, and $M_{\rm A}$, two of which are fixed ($\eta$ and $\varepsilon$) and the third one, $M_{\rm A}$, is running.

The basic equations and wave dispersion relations will be exposed in Sec.~2 of the paper, and the numerical results and discussions -- in Sec.~3.  Section 4 summarizes the new findings and comments on future improvements of this study.

\section{Basic equations and dispersion relations}
\label{sec:basiceqns}

The jet's interfaces are the surfaces $x = \pm x_0$, the uniform magnetic field $\mathbf{B}_0$ and the steady flow velocities $\mathbf{U}_{\rm o,e}$ point in the $z$ direction.  The wave vector $\mathbf{k}$ lies also along the $z$ axis and its direction is the same as that of $\mathbf{B}_0$ and $\mathbf{U}_{\rm o,e}$.  As we have already mentioned we will study the waves propagation in a reference frame affixed to the environment.  Thus the steady flow velocity in the slab is $\mathbf{U} \equiv \mathbf{U}^{\rm rel} = \mathbf{U}_{\rm o} - \mathbf{U}_{\rm e}$ and zero outside.  The basic equations for the incompressible Hall-MHD waves are the linearized equations governing the evolution of the perturbed fluid velocity $\delta \mathbf{v}$ and wave magnetic field $\delta \mathbf{B}$:
\begin{equation}
    \rho \frac{\partial}{\partial t} \delta \mathbf{v} + \rho(\mathbf{U} \cdot \nabla) \delta \mathbf{v} + \nabla(\frac{1}{\mu_0}\mathbf{B}_0 \cdot \delta \mathbf{B}) - \frac{1}{\mu_0} (\mathbf{B}_0 \cdot \nabla) \delta \mathbf{B} = 0,
\label{eq:movm}
\end{equation}
\begin{equation}
    \frac{\partial}{\partial t} \delta \mathbf{B} + (\mathbf{U} \cdot \nabla) \delta \mathbf{B} - (\mathbf{B}_0 \cdot \nabla) \delta \mathbf{v} + \mathbf{B}_0 \nabla \cdot \delta \mathbf{v} + \frac{v_{\rm A}^2}{\omega_{\rm ci}}\hat{z} \cdot \nabla \nabla \times \delta \mathbf{B} = 0,
\label{eq:magfield}
\end{equation}
with the constrains
\begin{equation}
    \nabla \cdot \delta \mathbf{v} = 0,
\label{eq:cont}
\end{equation}
\begin{equation}
    \nabla \cdot \delta \mathbf{B} = 0,
\label{eq:magind}
\end{equation}
where $v_{\rm A} = B_0/(\mu_0 \rho)^{1/2}$ is the Alfv\'en speed and $\mu_0$ is the permeability of free space.  After Fourier transforming all perturbed quantities 
$\propto\!\!g(x)\exp(-\mathrm{i}\omega t + \mathrm{i}kz)$, we derive two coupled second order differential equations for $\delta v_x$ and $\delta v_y$, notably
\begin{equation}
    \left( {\displaystyle \frac{\mathrm{d}^2}{\mathrm{d} x^2} - k^2} \right)\delta v_x - \mathrm{i} \frac{a - 1}{\epsilon} k^2 \delta v_y = 0,
\label{eq:short_1}
\end{equation}
and
\begin{equation}
    \left( {\displaystyle \frac{\mathrm{d}^2}{\mathrm{d} x^2} - k^2} \right)\delta v_x + \mathrm{i}{\displaystyle \frac{\epsilon}{a - 1}  \left( {\displaystyle \frac{\mathrm{d}^2}{\mathrm{d} x^2} - k^2} \right)} \delta v_y = 0,
\label{eq:short_2}
\end{equation}
where
\[
    \epsilon = \frac{\omega - \mathbf{k} \cdot \mathbf{U}}{\omega_{\rm ci}}, \quad \mbox{and} \quad a = \left( \frac{\omega - \mathbf{k} \cdot \mathbf{U}}{kv_{\rm A}} \right)^2
\]
with $\omega_{\rm ci} = eB_0/m_{\rm i}$ being the ion angular cyclotron frequency ($m_{\rm i}$ is ion/proton mass and $e$ is the elementary electric charge).  Note that we have different $\epsilon$s and $a$s inside and outside the jet.  We seek the solutions to coupled equations (\ref{eq:short_1}) and (\ref{eq:short_2}) in the form
\[
    \delta v_x(x) = f \left[ \exp(-\kappa x) \mp \exp(\kappa x)
    \right],
\]
\[
    \delta v_y(x) = \mathrm{i} h \left[ \exp(-\kappa x) \mp \exp(\kappa x)
    \right],
\]
anticipating surface waves with attenuation coefficient $\kappa$, and obtain the following set of equations
\begin{eqnarray*}
    \left( \kappa^2 - k^2 \right)f + \frac{a - 1}{\epsilon}
    k^2h = 0, \\
    \left( \kappa^2 - k^2 \right)f - \frac{\epsilon}{a - 1}
    \left( \kappa^2 - k^2 \right)h = 0.
\end{eqnarray*}
This set of equations yields the following expressions for $\kappa$:
\begin{eqnarray}
    \kappa_1 &=& k, \nonumber
    \\
    \kappa_2 = k\left[ 1 - (a - 1)^2/\epsilon^2 \right]^{1/2} &\equiv& m. \nonumber
\label{eq:kappa}
\end{eqnarray}
That means that there are in fact two pairs of attenuation coefficients: ($k,m_{\rm o}$) inside the flux tube and ($k,m_{\rm e}$) outside the jet, respectively.

As is known, on a bounded MHD plasma wave guide (cylinder or layer) two types of waves may exist.  Recall that for a slab geometry the general solutions to the equations governing $\delta v_x$ and $\delta v_y$ are sought in the form of superpositions of $\cosh \kappa_{\rm o}x$ and $\sinh \kappa_{\rm o}x$.  Those
solutions contain waves whose shape is defined by the $\cosh$ function (they are called \emph{kink\/} waves) and another type of waves associated with the $\sinh$ function (which are termed \emph{sausage\/} waves).  The transverse structure of both waves inside the slab is determined by the two attenuation coefficients $\kappa_{\rm o1,2}$ (i.e., $k$ and $m_{\rm o}$).  Thus the solutions for $\delta
v_x$ and $\delta v_y$ inside the slab ($|x| < x_0$) assuming a sausage wave form, accordingly, are
\[
    \delta v_x(x) = f_1 {\displaystyle \frac{\sinh\,k x} {\sinh\,k x_0}}
      + f_2 {\displaystyle \frac{\sinh\,m_{\rm o} x}
            {\sinh\,m_{\rm o} x_0}},
\]
and
\[
     \delta v_y(x) = \mathrm{i} f_1 G_{\rm o1} \displaystyle
            {\frac{\sinh\,k x} {\sinh\,k x_0}}
     + \mathrm{i} f_2 G_{\rm o2} \displaystyle {\frac{\sinh\,m_{\rm o} x}
            {\sinh\,m_{\rm o} x_0}},
\]
where
\[
    G_{\rm o1,2} = \displaystyle{
    -\frac{\epsilon_{\rm o}}{a_{\rm o} - 1} \frac{\kappa_{\rm o1,2}^2 - k^2}{k^2}}.
\]
These solutions have been obtained by imposing the compatibility condition for the set of equations for $f$ and $h$ as unknown quantities.
We also notice that $G_{\rm o1} = 0$ and $G_{\rm o2} = (a_{\rm o} - 1)/\epsilon_{\rm o}$.  For a kink surface-wave form the expressions for perturbed fluid velocity components have the same description -- it is only
necessary to replace $\sinh$ with $\cosh$.  The solutions outside the layer (identical for both modes) are
\[
    \delta v_x(x) = \begin{cases}
     \alpha_1 \exp \left[ -k(x - x_0) \right] + \alpha_2 \exp \left[ -m_{\rm e}(x - x_0) \right] & \text{for $x > x_0$,}
     \\
     \beta_1 \exp \left[ k(x + x_0) \right] + \beta_2 \exp \left[ m_{\rm e}(x + x_0) \right] & \text{for $x < -x_0$,} \end{cases}
\]
and
\[
     \delta v_y(x) = \begin{cases}
      \mathrm{i} \alpha_1 G_{{\rm e}1} \exp\left[ -k(x - x_0) \right] + \mathrm{i}
      \alpha_2 G_{{\rm e}2} \exp \left[ -m_{\rm e}(x - x_0) \right] &
      \text{for $x > x_0$,}
      \\
      \mathrm{i} \beta_1 G_{{\rm e}1} \exp \left[ k(x + x_0) \right] + \mathrm{i}
      \beta_2 G_{{\rm e}2} \exp \left[ m_{\rm e}(x + x_0) \right] &
      \text{for $x < -x_0$.} \end{cases}
\]
Here, as above,
\[
    G_{\rm e1,2} = \displaystyle{
    -\frac{\epsilon_{\rm e}}{a_{\rm e} - 1} \frac{\kappa_{\rm e1,2}^2 - k^2}{k^2}},
\]
and, similarly, $G_{\rm e1} = 0$ and $G_{\rm e2} = (a_{\rm e} - 1)/\epsilon_{\rm e}$.

Having derived the expressions for the perturbed fluid velocity components $\delta v_x$ and $\delta v_y$, one can calculate the perturbed total pressure, which in our case is the perturbed magnetic pressure only, and arrive at
\begin{eqnarray*}
     \delta p_{\rm total}(x) = \frac{1}{\mu_0}\mathbf{B}_0
     \cdot \delta \mathbf{B}
     {}= \mathrm{i} \frac{\rho}{\omega - \mathbf{k} \cdot \mathbf{U}} v_{\rm
     A}^2 \\ \\
     \times\left\{ {\displaystyle
     \left[ \frac{\left( \omega - \mathbf{k} \cdot \mathbf{U}
     \right)^2}{k^2 v_{\rm A}^2} - 1 \right] \frac{\mathrm{d}}{\mathrm{d} x}
     \delta v_x}(x) +
     \mathrm{i}\:{\displaystyle \frac{\omega - \mathbf{k} \cdot \mathbf{U}}
     {\omega_{\rm ci}} \frac{\mathrm{d}}{\mathrm{d} x} \delta v_y(x)}
     \right\}.
\end{eqnarray*}
\vspace{1mm}

The perturbed wave electric field $\delta \mathbf{E}$ can be obtained from the generalized Ohm's law
\[
     \mathbf{E} = - \mathbf{v} \times \mathbf{B} +
       \frac{m_{\rm i}}{e \rho}\, \mathbf{j} \times \mathbf{B},
\]
which, by means of Amp\`ere's law (multiplied vectorially by $\mathbf{B}$) yields
\[
     \delta \mathbf{E} = \mathbf{B}_0 \times
     \left( \delta \mathbf{v}
     - {\displaystyle l_{\rm Hall} \frac{v_{\rm A}}{B_0}}
     \nabla \times \delta
     \mathbf{B} \right) - \mathbf{U} \times \delta \mathbf{B}.
\]
Its components are
\[
    \delta E_x(x) = -{\displaystyle \frac{\omega}{\omega -
    \mathbf{k} \cdot \mathbf{U}} \left( \frac{\omega - \mathbf{k}
    \cdot \mathbf{U}}{k v_{\rm A}} \right)^2} B_0 \delta v_y(x),
\]
and
\[
    \delta E_y(x) = {\displaystyle \frac{\omega}{\omega -
    \mathbf{k} \cdot \mathbf{U}} B_0 \left[ \delta v_x(x) -
    \mathrm{i}\: \frac{\omega - \mathbf{k} \cdot \mathbf{U}}{\omega_{\rm ci}}
    \delta v_y(x) \right]}.
\]
The above expressions for perturbed quantities are used when
implementing the boundary conditions.

It follows from the solutions to the basic equations (\ref{eq:short_1}) and (\ref{eq:short_2}) for the perturbed fluid velocity components $\delta v_x$ and $\delta v_y$ that the number of integration constants is six.  However, because of the symmetry (or antisymmetry), the two $\beta_{1,2}$ amplitudes are in fact
directly obtainable from the $x > x_0$ solutions -- indexed $\alpha$s and $\beta$s are not independent.  Thus we can derive the dispersion relations by applying only four boundary conditions at one interface, for example, at $x = x_0$.  These boundary conditions, as we already mentioned in Sec.~\ref{sec:intro}, are derived and discussed in \cite{ivan96}.  We can borrow them except the first one, the continuity of $\delta v_x$ across the interface, as in the present case of a flowing slab this condition must be replaced by the continuity of $\delta v_x/\left(\omega - \mathbf{k} \cdot \mathbf{U} \right)$ \cite{chandra61}.  The rest of the boundary conditions are the continuity of the  perturbed pressure $\delta p_{\rm total}$, the $y$-component of perturbed wave electric field $\delta E_y$, and the $x$-component of perturbed electric displacement $\delta D_x = \varepsilon_0 \left( K_{xx}\delta E_x + K_{xy}\delta E_y \right)$ (where $\varepsilon_0$ is the permittivity of the free space) at the interface.  In the last boundary condition, $K_{xx}$ and $K_{xy}$ are the low-frequency components of the plasma dielectric tensor \cite{swanson89}
\[
     K_{xx} \approx \frac{c^2}{v_{\rm A}^2} \qquad {\rm and}
     \qquad K_{xy} \cong \mathrm{i} \frac{\omega -
     \mathbf{k} \cdot \mathbf{U}}{\omega_{\rm ci}} \frac{c^2}{v_{\rm A}^2}.
\]

By imposing the boundary conditions at the interface $x = x_0$, and after some straightforward algebra, we finally arrive at the dispersion relations for parallel propagation of sausage and kink waves in a planar jet, surrounded by steady plasma media
\begin{eqnarray}
\label{eq:dispeqn}
     \left( \frac{\omega - \mathbf{k} \cdot \mathbf{U}}
     {k v_{\rm Ao}}\right)^2 - 1 + \left[ \frac{\rho_{\rm e}}{\rho_{\rm o}}
     \left( \frac{\omega}{k v_{\rm Ao}}\right)^2 - 1 \right] {\tanh \choose \coth}kx_0
     \nonumber \\
     \\ \nonumber
     {}- \epsilon_{\rm o}^2 \left[ 1 + \tilde{\omega}^2 \frac{\rho_{\rm e}}{\rho_{\rm o}}
     {\tanh \choose \coth}kx_0 \right]\frac{1 - \tilde{\omega}
     \rho_{\rm e}/\rho_{\rm o}}{1 - \tilde{\omega}
     \left( \rho_{\rm e}/\rho_{\rm o} \right)^2 } = 0,
\end{eqnarray}
where
\[
    \tilde{\omega} = \frac{\omega}{\omega - \mathbf{k} \cdot \mathbf{U}} \quad \mbox{and} \quad \epsilon_{\rm o} = \frac{\omega - \mathbf{k}\cdot \mathbf{U}}
    {\omega_{\rm ci}}.
\]
As can be seen, the wave frequency $\omega$ is Doppler-shifted inside the jet. These dispersion relations can be extracted from Eq.~(12) in \cite{rossi04} in the limit $c_{\rm s} \to \infty$ (sound speed much larger than the Alfv\'en one) by letting $\mathbf{U}_{\rm e} = 0$ and $\mathbf{U}_{\rm o} = \mathbf{U}$.  When the flux tube is static ($\mathbf{U} = 0$), Eq.~(\ref{eq:dispeqn}) coincides with Eq.~(18) in Ref.~\cite{ivan96}.  Finally, when one neglects the Hall effect ($\epsilon_{\rm o} = 0$) one gets the well-known, derived by Edwin and Roberts \cite{pat82}, dispersion relations in the limit $c_{\rm s} \to \infty$.  We should also emphasize that in an incompressible jet both modes are pure surface waves.  A question which immediately arises is what type of MHD waves are those described by Eq.~(\ref{eq:dispeqn})?  From the three well-known MHD linear modes propagating in infinite compressible magnetized plasmas (Alfv\'en, fast and slow magnetosonic waves) in incompressible limit survive only two, notably the shear and pseudo Alfv\'en waves.  The latter is the incompressible vestige of the slow mode of compressible MHD.  The displacement vector of a shear Alfv\'en wave is perpendicular to the plane defined by its wave vector, $\mathbf{k}$, and a uniform background magnetic field, $\mathbf{B}_0$, whereas that of a pseudo Alfv\'en wave lies in this plane. The two wave modes share the dispersion relation
\[
    \omega^2 = \frac{(\mathbf{k} \cdot \mathbf{B}_0)^2}{\mu_0 \rho} \equiv \left( k_{\parallel}v_{\rm A} \right)^2
\]
and propagate with group velocity, $\mathbf{v}_{\rm A}$, either parallel or antiparallel to $\mathbf{B}_0$ depending upon the sign of $k_\parallel$.  It seems that our Hall-MHD modes travelling along the jet are akin to the pseudo Alfv\'en waves.

The dispersion relation of each mode can be symbolically written down in the form:
\begin{equation}
    \mathcal{D}(\omega,k,\mathrm{parameters}) = 0,
\label{eq:symb}
\end{equation}
where the function argument `parameters' includes data specific for the jet -- they will be listed shortly.  As we are interested in the stability of the surface waves running at the jet interfaces, we have to assume that the wave frequency is complex, i.e., $\omega \to \omega + \mathrm{i}\gamma$, where $\gamma$ is the expected instability growth rate.  Thus dispersion equations (\ref{eq:dispeqn}) become complex and their solving is not a trivial problem \cite{acton90}.  When studying dispersion characteristics of MHD waves, one usually plots the dependence of the wave phase velocity $v_{\rm ph}$ as function of the wave number $k$.  For numerical solving of equations (\ref{eq:dispeqn}) we normalize all quantities by defining the dimensionless wave phase velocity $V_{\rm ph} = \omega/kv_{\rm Ao}$, wave number $K = kx_0$, and the (relative) Alfv\'enic Mach number $M_{\rm A} = U/v_{\rm Ao}$, respectively.  Accordingly
\[
    \left[ (\omega - \mathbf{k} \cdot \mathbf{U})/(kv_{\rm Ao}) \right]^2 = \left( V_{\rm ph} - M_{\rm A} \right)^2, \quad  \omega^2/(kv_{\rm Ao})^2 = V_{\rm ph}^2,
\]
\[
    \tilde{\omega} = V_{\rm ph}/
    \left( V_{\rm ph} - M_{\rm A} \right) \quad \mbox{and} \quad
    \epsilon_{\rm o} = K \left( V_{\rm ph} - M_{\rm A} \right)
    l_{\rm Hall}/x_0.
\]
Note, that $l_{\rm Hall}/x_0 = \varepsilon$ (alongside with $\eta$ and $M_{\rm A}$) is an entry parameter which has to be specified at the start of the numerical procedure.  Thus we have to solve normalized dispersion relations (\ref{eq:dispeqn}), having now the form
\begin{eqnarray}
\label{eq:numdisp}
    \left( V_{\rm ph} - M_{\rm A} \right)^2 - 1 + \left( \eta V_{\rm ph}^2 - 1 \right){\tanh \choose \coth}K  \nonumber \\
    \\ \nonumber
    {}- K^2 \varepsilon^2 \left[ \left( V_{\rm ph} - M_{\rm A} \right)^2 + V_{\rm ph}^2 \eta {\tanh \choose \coth}K \right]\frac{V_{\rm ph}(1 - \eta) - M_{\rm A}} {V_{\rm ph}(1 - \eta^2) - M_{\rm A}} = 0.
\end{eqnarray}
Recall that we consider the normalized wave phase velocity $V_{\rm ph}$ as a complex number and we shall look for the dependencies of the real and imaginary parts of $V_{\rm ph}$ as functions of the real dimensionless wave number $K$ at given values of the three entry parameters.  It can be easily seen from above equations that they are cubic ones with respect to $V_{\rm ph}$.  Hence, the dimensionless dispersion relation of, for example, the kink mode can be displayed in the form:
\begin{equation}
    A\,V_{\rm ph}^3 + B\,V_{\rm ph}^2 + C\,V_{\rm ph} + D = 0,
\label{eq:kink}
\end{equation}
where
\begin{eqnarray*}
    A &=& (1 + \eta \coth K)(1 - \eta)(1 + \eta - \varepsilon^2 K^2),\\
    B &=& -M_{\rm A}\left[ 2(1 - \eta)(1 + \eta - \varepsilon^2 K^2) + (1 + \eta \coth K)(1 - \varepsilon^2 K^2) \right],\\
    C &=& M_{\rm A}^2 \left[ (1 - \eta)(1 + \eta - \varepsilon^2 K^2) + 1 - \varepsilon^2 K^2 \right] - (1 + \coth K)(1 - \eta^2),\\
    D &=& M_{\rm A} \left[ -M_{\rm A}^2 (1 - \varepsilon^2 K^2) + 1 + \coth K   \right].
\end{eqnarray*}
The dimensionless dispersion relation for sausage mode is similar -- one has simply to replace $\coth$ by $\tanh$.

Cubic equations will be solved on using Cardano's formulas \cite{kurosh65}.  First let us define a variable $f$:
\[
    f =  -\frac{1}{3}\frac{B^2}{A^2} + \frac{C}{A}.
\]
Next we define $g$:
\[
    g = \frac{2}{27}\frac{B^3}{A^3} - \frac{BC}{3A^2} + \frac{D}{A}.
\]
Finally we define $h$:
\begin{equation}
    h = (f/3)^3 + (g/2)^2.
\label{eq:h}
\end{equation}
If $h > 0$, there is only one real root and two complex conjugate ones.  When $h \leqslant 0$, all three roots are real.

When $h < 0$, the real solutions to the cubic Eq.~(\ref{eq:kink}) are:
\begin{equation}
\label{eq:v1}
    V_1 = 2\sqrt{-\frac{f}{3}}\cos \frac{\alpha}{3} - \frac{B}{3A},
\end{equation}
\begin{equation}
\label{eq:v23}
    V_{2,3} = -2\sqrt{-\frac{f}{3}}\cos \left( \frac{\alpha}{3} \pm \frac{\pi}{3} \right) - \frac{B}{3A},
\end{equation}
where
\[
    \alpha = \cos^{-1}\left[-\frac{g}{2\sqrt{-(f/3)^3}}\right].
\]
For simplicity, we have dropped the subscript `ph' to the normalized phase velocity $V$.

When $h > 0$, the real root of Eq.~(\ref{eq:kink}) is given by
\begin{equation}
\label{eq:v0}
    V_0 = M + N - \frac{B}{3A},
\end{equation}
where
\[
    M = \left( -g/2 + \sqrt{h} \right)^{1/3} \quad \mbox{and} \quad N = \left( -g/2 - \sqrt{h} \right)^{1/3}.
\]
The two complex roots are given by $V_{\rm r} \pm \mathrm{i}V_{\rm i}$, where
\begin{equation}
\label{eq:Vr}
    V_{\rm r} = -\frac{1}{2}(M + N) - \frac{B}{3A},
\end{equation}
and
\begin{equation}
\label{eq:Vi}
    V_{\rm i} = \frac{\sqrt{3}}{2}(M - N).
\end{equation}

During the numerical solving of dispersion equations (\ref{eq:numdisp}) we can look what sign possesses $h$, defined by Eq.~(\ref{eq:h}), and in the ranges of the normalized wave numbers $K$, where $h$ is positive, we can expect complex solutions, i.e., amplification or damping of the waves due to their interaction with the flow.

\section{Numerical results and discussion}
\label{sec:results}

Before starting the numerical solving of dispersion equations (\ref{eq:numdisp}) we have to specify the entry jet's parameters [c.f., Eq.~(\ref{eq:symb})].  As was told in the Introduction section, we take the Hall parameter $\varepsilon = 0.4$.  The relative Alfv\'enic Mach number, $M_{\rm A}$, will be a running parameter, from $0$ (for a static flux tube) to some reasonable values.  These values to some extent depend upon the choice of the third parameter, $\eta$, equal to the ratio of plasma densities outside and inside the jet.  When studying Kelvin--Helmholtz instabilities of MHD waves in the coronal plume--interplume region in the framework of standard magnetohydrodynamics, Andries and Goossens \cite{andries01} find that in the $\beta = 0$ case one can expect that the instability will occur approximately for $|U| > v_{\rm Ao} + v_{\rm Ae}$.  As we have already mentioned in Sec.~\ref{sec:intro}, we assume that this estimation is valid for our case of an incompressible plasma jet.  After normalizing all velocities with respect to the Alfv\'en speed inside the slab, $v_{\rm Ao}$, we get
\[
    \frac{|U|}{v_{\rm Ao}} > 1 + \frac{v_{\rm Ae}}{v_{\rm Ao}}.
\]
or
\begin{equation}
\label{eq:eta}
    \left| M_{\rm A} \right| > 1 + \frac{1}{\sqrt{\eta}}.
\end{equation}
As seen from above inequality, for some small $\eta$s the relative Alfv\'enic Mach number, $M_{\rm A}$, might become rather large in order to register the onset of Kelvin--Helmholtz instability.  Such large Mach numbers imply that for magnitudes of the Alfv\'en speed inside a solar wind jet in the range, say, of $60$--$100$ km\,s$^{-1}$ the difference between Alfv\'en speeds in the jet and its environment should be of the order of a few hundred kilometers per second which is unlikely to occur.  Moreover, for relatively large Alfv\'enic Mach numbers dispersion curves of Hall-MHD surface modes become rather complicated.  That is why we choose $\eta = 10$, which means that one can expect a onset of instability at $\left| M_{\rm A} \right| > 1.316$ that looks as a reasonable value.

A specific feature of the surface Hall-MHD waves travelling along an incompressible static plasma layer is that there exists a limiting dimensionless wave number $K_{\rm limit}$ beyond which the wave propagation is no longer possible.  That limiting wave number is given by \cite{ivan96}
\begin{equation}
\label{eq:klimit}
    K_{\rm limit} = (1 + \eta)^{1/2}/\varepsilon.
\end{equation}
For our choice of $\eta$ and $\varepsilon$, $K_{\rm limit} = 8.292$.  With approaching that wave number the wave phase velocity becomes very large.  It is interesting to see whether the steady flow will change that limiting value.

\begin{figure}[!ht]
\centering\includegraphics[height=.30\textheight]{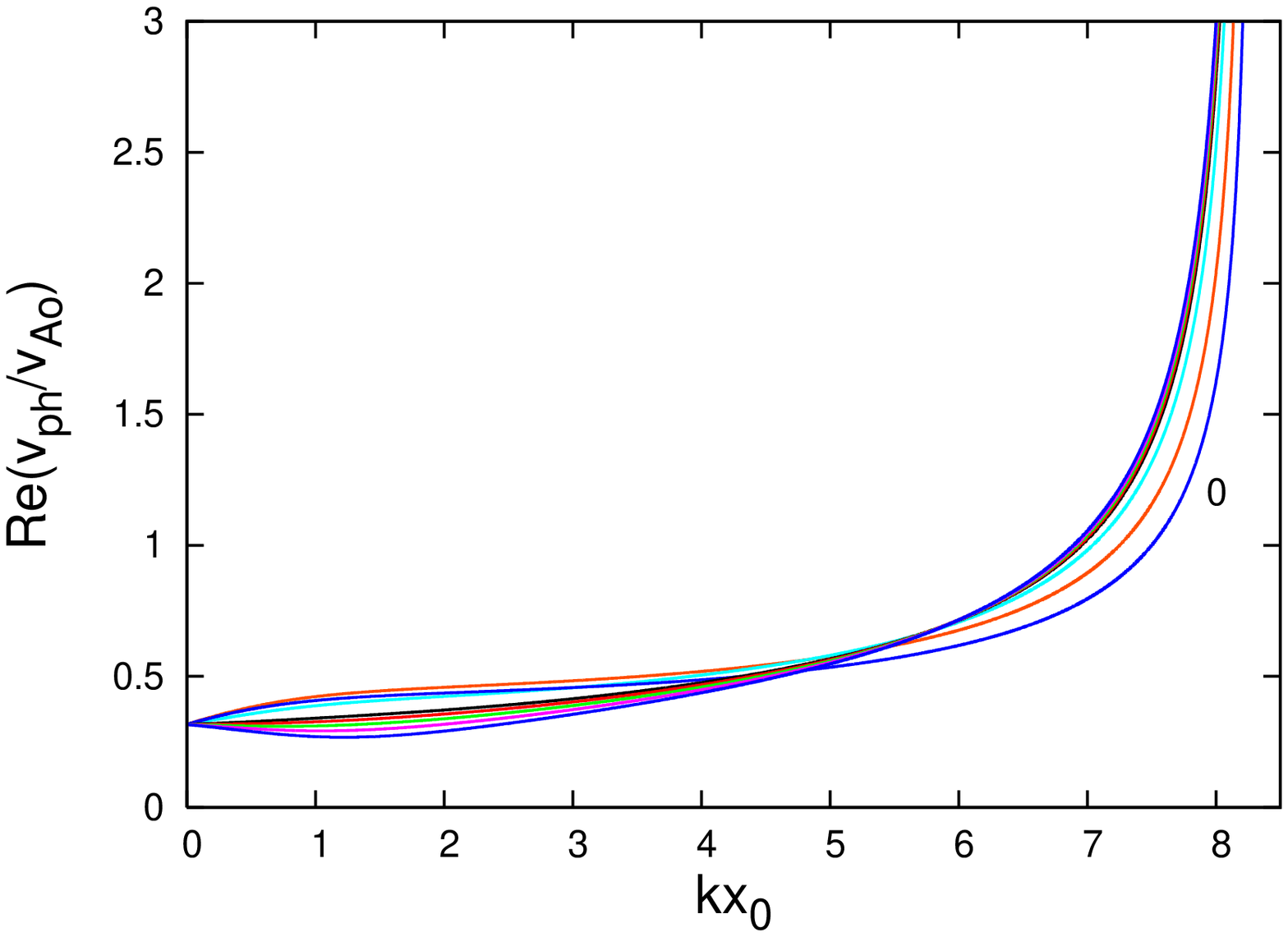}
  \caption{(Online colour) Dispersion curves of kink Hall-MHD waves travelling along an incompressible flowing plasma layer for positive values of the relative Alfv\'enic Mach number $M_{\rm A}$ -- all waves are stable.  For details in curves' labeling see Fig.~\ref{fig:fg2}.
  \label{fig:fg1}}       
\end{figure}

\begin{figure}[!hb]
\centering\includegraphics[height=.30\textheight]{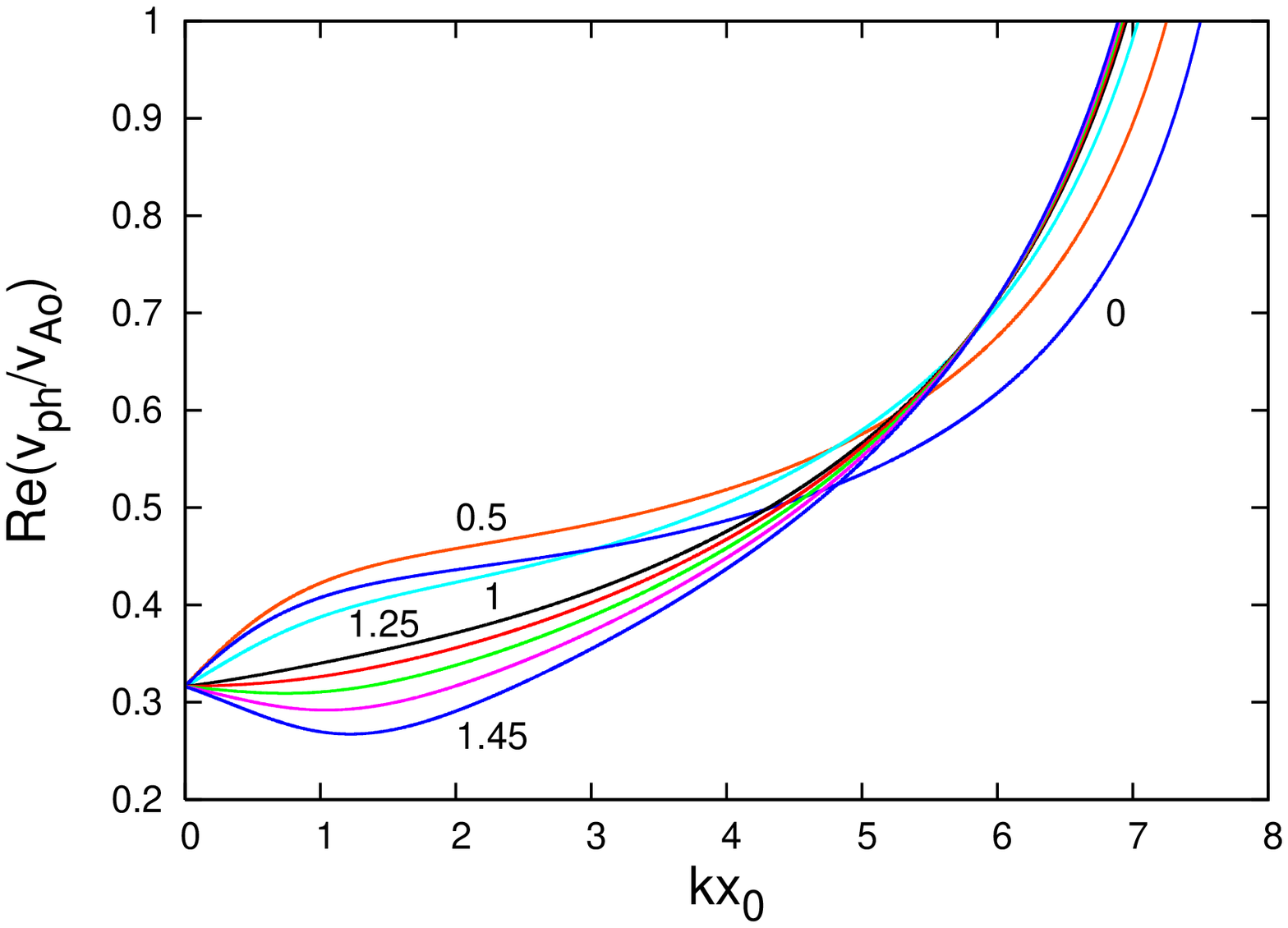}
  \caption{(Online colour) A zoom of the bottom part of Fig.~\ref{fig:fg1}.  The dispersion curves sandwiched between curves labeled by `1.25' and `1.45' correspond to $M_{\rm A}$ equal to $1.3$, $1.35$, and $1.4$, respectively.
  \label{fig:fg2}}       
\end{figure}

Let us first start with the kink mode.  For solving the corresponding dispersion relation (\ref{eq:numdisp}) we generally use the roots of the cubic equation given by Eqs.~(\ref{eq:v1})--(\ref{eq:Vi}).  We begin the numerical solving with the relative $M_{\rm A} = 0$ (corresponding to a static flux tube) running the dimensionless wave number $K$ from $0.005$ to $10$.  As naturally to expect, in that wave number region $h$ is negative and we get real values for the normalized phase velocity -- the corresponding dispersion curve is labeled by `0' in Fig.~\ref{fig:fg1}.  As seen from that figure, the kink wave is generally a sub-Alfv\'enic one; at $K \approx 7.2$ its phase velocity becomes equal to the Alfv\'en speed and starts quickly to grow up reaching rather large values (up to $800$ even more) whence $K \to K_{\rm limit}$.  What is going on as the layer possesses any flow velocity?  If we take, for instance, $M_{\rm A} = 0.5$, one can construct from the roots of the cubic equation two dispersion curves, both being real solutions, however, one curve with a positive phase velocity, and another curve with a negative one.  The first dispersion curve as seen from Fig.~\ref{fig:fg2} is slightly above the `$0$'th curve, while the second dispersion curve starts with the negative value of $-0.32$ which becomes large in magnitude at $K$ approaching its critical value, and what is more interesting, it goes beyond the $K_{\rm limit}$, now decreasing in magnitude. In this extended propagation range there exists a complex solution to the cubic equation with positive imaginary part, i.e., the wave becomes unstable.  However, bearing in mind that it is unlikely to observe/detect such a backward wave in a solar wind tube (supposing that the wave phase and group velocities have the same direction), we have to drop all the solutions with negative phase velocities.  The dispersion curves of the second type will be discussed in another, physically acceptable situation, soon.

With increasing the relative Alfv\'enic Mach number the dispersion curves with positive velocities initially lie below the `$0$'th curve, but afterwards, for values of the normalized wave number between $3$ and $5.5$, they cross the curve with label `0' staying on the left side of that curve.  In other words, the flow velocity slightly diminishes the value of the $K_{\rm limit}$ without allowing them to propagate beyond those limiting $K$s.  Note also that the course of the dispersion curves is not monotonous -- initially, for small $M_{\rm A}$s, they lie above the neutral (`$0$'th labeled) curve in the range of small and average dimensionless wave numbers, while for $M_{\rm A} \geqslant 1$ those curves set up below neutral dispersion curve.

\begin{figure}[!ht]
\centering\includegraphics[height=.30\textheight]{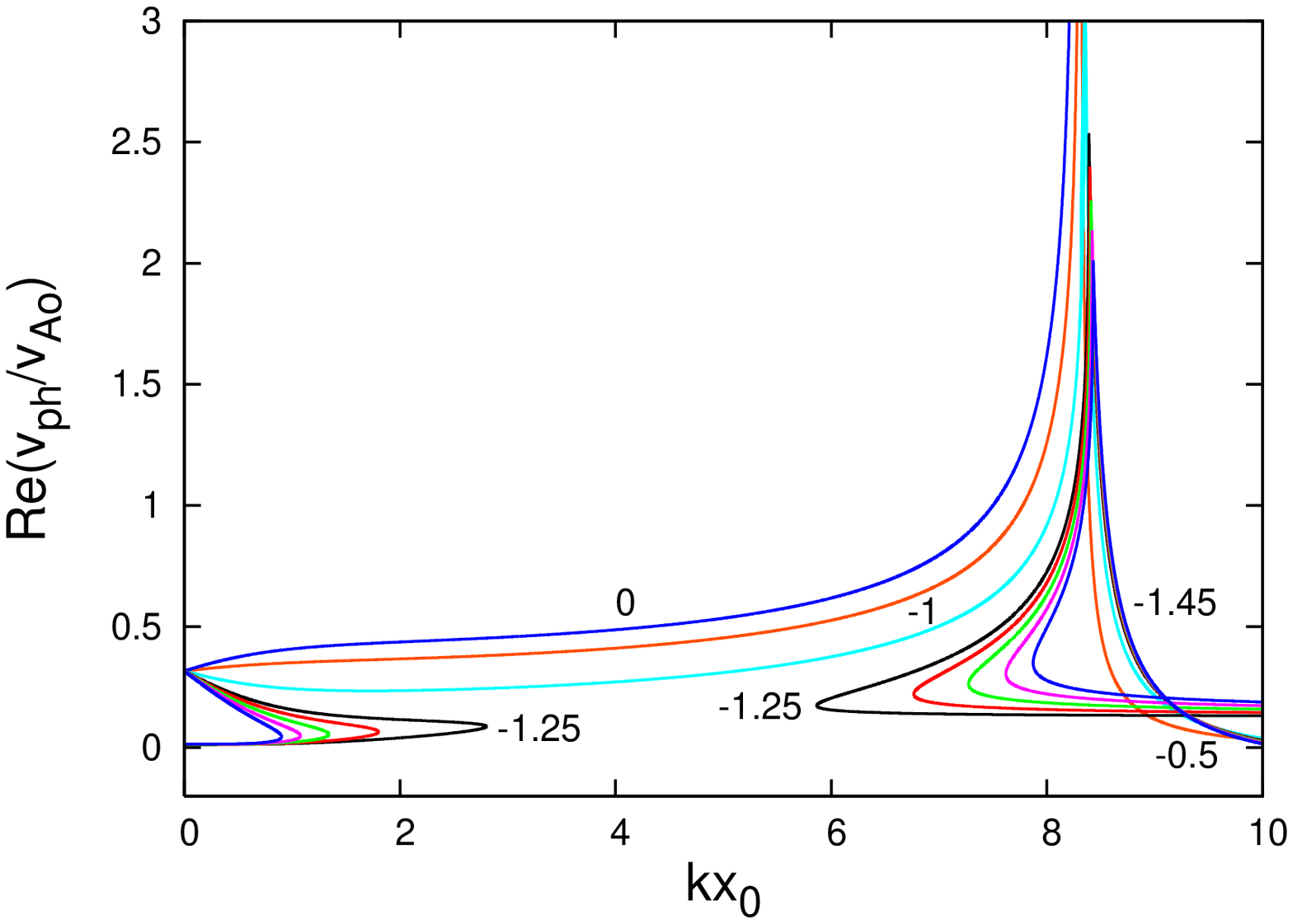}
  \caption{(Online colour) Dispersion curves of kink Hall-MHD waves travelling along an incompressible flowing plasma layer for negative values of the relative Alfv\'enic Mach number $M_{\rm A}$.  All the curves lying on the left side of the vertical merging line/cusp represent stable waves' propagation.  The waves become unstable only on the family of curves labeled by `$-0.5$' and `$-1.45$'.  For details in curves' labeling see Fig.~\ref{fig:fg4}.
  \label{fig:fg3}}       
\end{figure}

\begin{figure}[!hb]
\centering\includegraphics[height=.30\textheight]{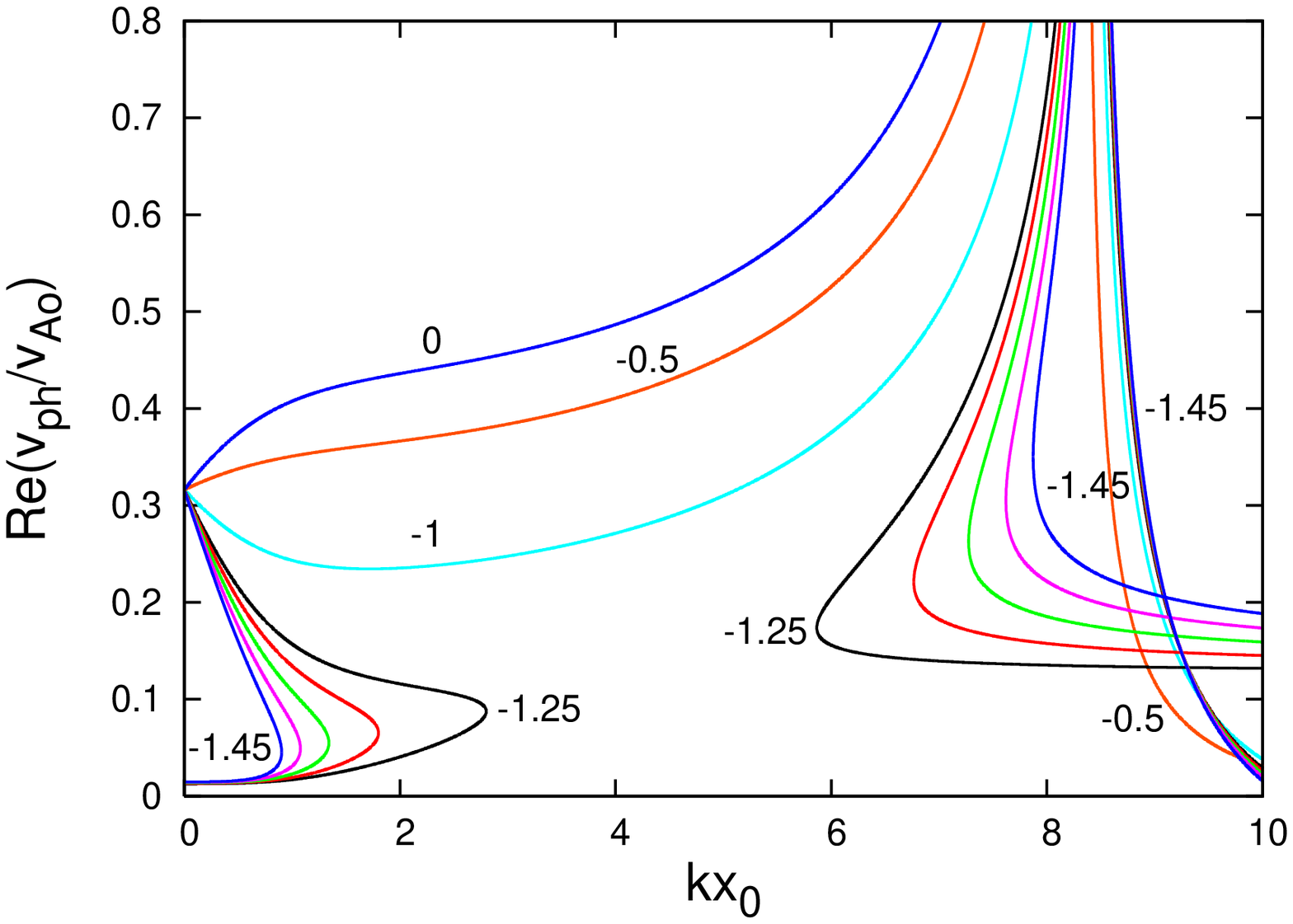}
  \caption{(Online colour) A zoom of the bottom part of Fig.~\ref{fig:fg3}.  The dispersion curves situated between the curves labeled by `$-1.25$' and `$-1.45$' correspond to $M_{\rm A}$ equal to $-1.3$, $-1.35$, and $-1.4$, respectively.
  \label{fig:fg4}}       
\end{figure}

\begin{figure}[!ht]
\centering\includegraphics[height=.30\textheight]{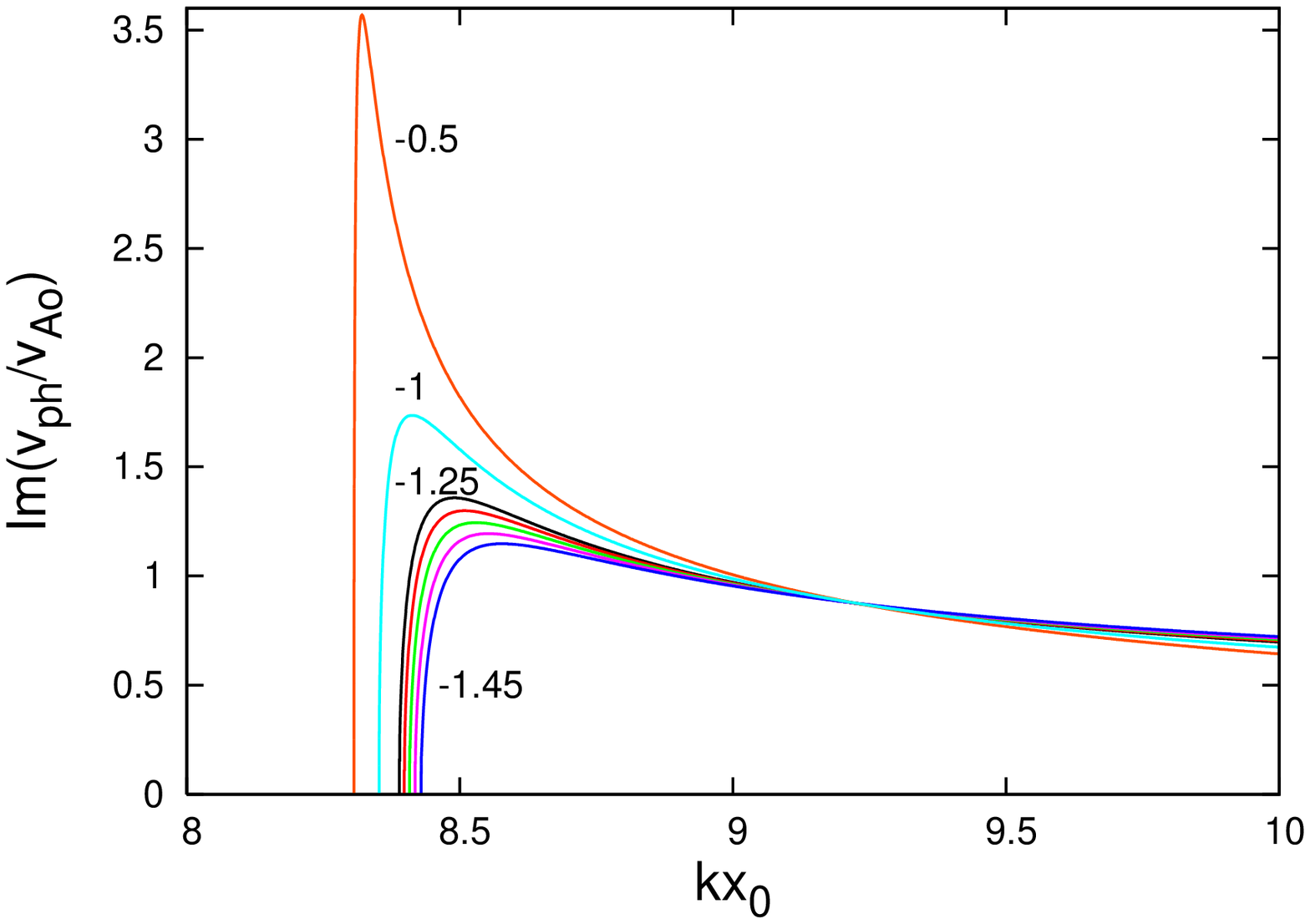}
  \caption{(Online colour) Growth rates of unstable kink Hall-MHD waves travelling along an incompressible flowing plasma layer for negative values of the relative Alfv\'enic Mach number $M_{\rm A}$ in the short-wavelength region (beyond the $K_{\rm limit}$).
  \label{fig:fg5}}       
\end{figure}

For any negative value of the relative Alfv\'enic Mach number $M_{\rm A}$ we have as before two set of solutions.  The real ones are negative and represent a mirror image of dispersion curves shown in Fig.~\ref{fig:fg1}.  We do not plot them for the same reason; although mathematically correct they are not acceptable from a physical point of view.  The most interesting case are the dispersion curves shown in Figs.~\ref{fig:fg3} and \ref{fig:fg4}.  It is clearly seen that for each $M_{\rm A}$, in fact, two distinctive dispersion curves merge at $K \approx K_{\rm limit}$.  The solutions to the dispersion relation for the curves lying on the right side of the merging vertical line at some $K$s become complex with positive imaginary part, i.e., there the waves are unstable.  The other important observation is that now the waves' phase velocities do not grow too much when $K \to K_{\rm limit}$.  That is especially true for the dispersion curves associated with relatively large in modulus $M_{\rm A}$s -- see, for example, the dispersion curve labeled by `$-1.45$.'  Another important feature is the circumstance that all dispersion curves lying on the left side of the merging vertical line correspond to a stable (generally with complicated shapes of the dispersion curves) waves' propagation!  We also note that parts of the dispersion curves, corresponding to a stable wave propagation, continue smoothly beyond the $K_{\rm limit}$ -- see, for example, in Fig.~\ref{fig:fg4} the dispersion curve labeled by `$-1.25$' that ends at $kx_0 = 10$ with $v_{\rm ph}/v_{\rm Ao} = 0.13$.  It turns out that the Hall current makes the waves stable in the wave number propagation range between $0$ and $K_{\rm limit}$ -- an instability onset is only possible at some critical wave numbers larger than $K_{\rm limit}$.  The growth rates of kink waves in the instability region are plotted in Fig.~\ref{fig:fg5}.  We would like to emphasize that all complex solutions were checked for some selected dimensionless wave numbers on using the complex versions of Newton--Raphson and M\"{u}ller \cite{muller56} methods.

\begin{figure}[!hb]
\centering\includegraphics[height=.30\textheight]{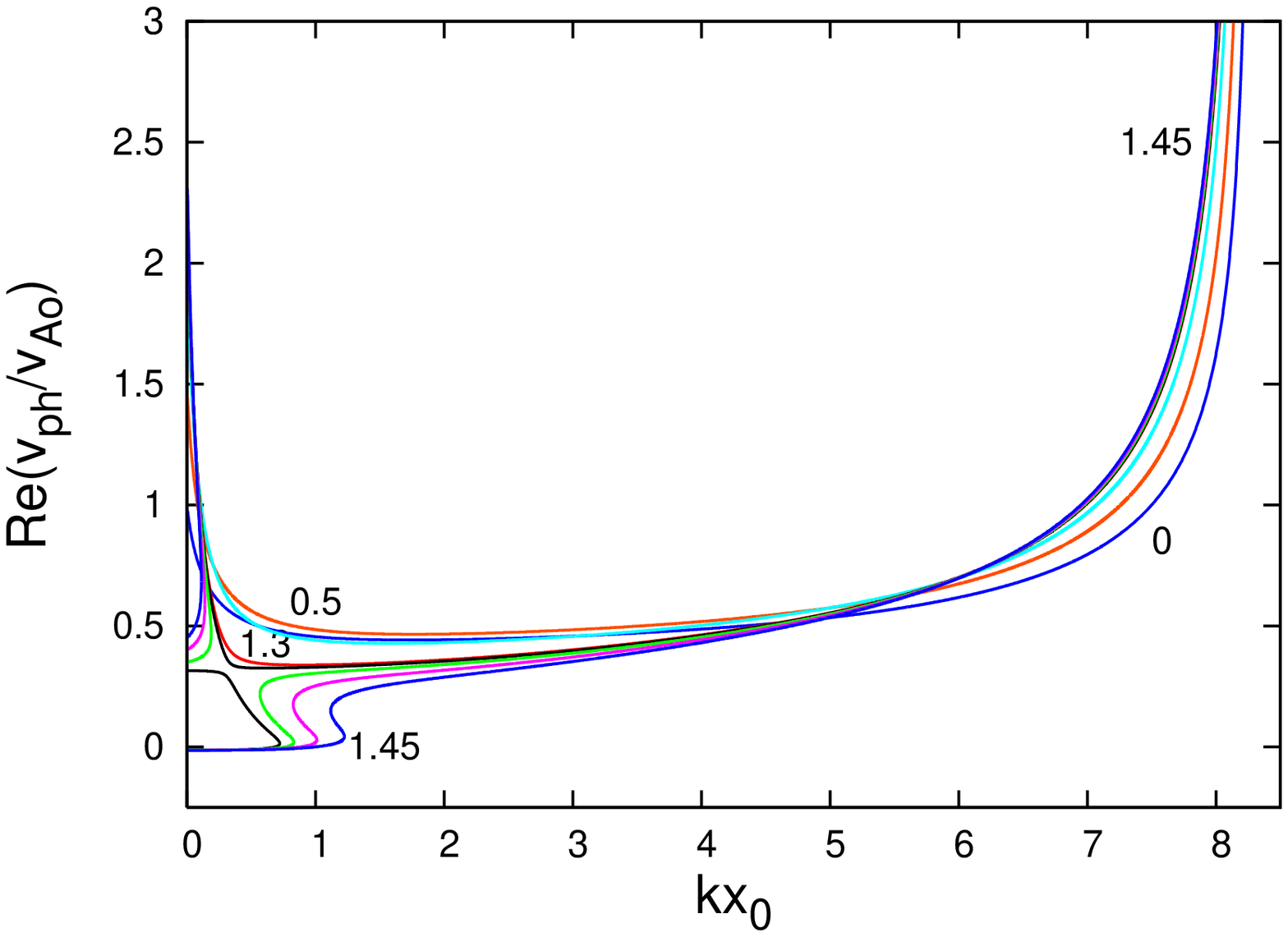}
  \caption{(Online colour) Dispersion curves of sausage Hall-MHD waves travelling along an incompressible flowing plasma layer for positive values of the relative Alfv\'enic Mach number $M_{\rm A}$ -- all waves are stable.  For details in curves' labeling see Fig.~\ref{fig:fg7}.
  \label{fig:fg6}}       
\end{figure}

\begin{figure}[!ht]
\centering\includegraphics[height=.30\textheight]{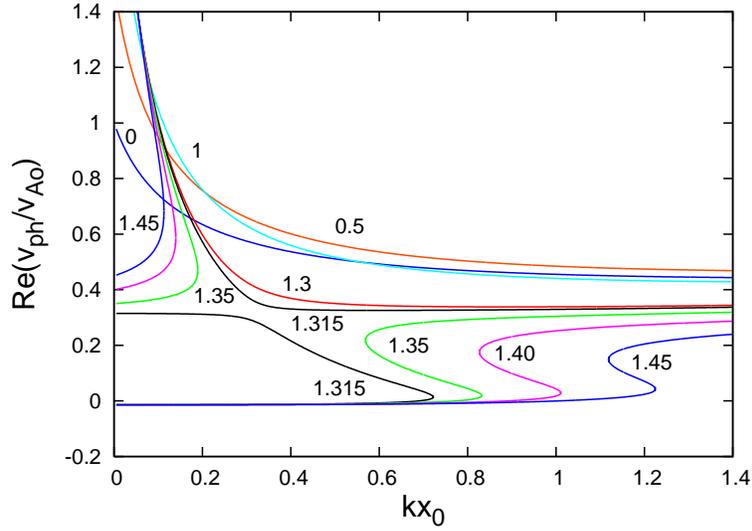}
  \caption{(Online colour) A zoom of the bottom part of Fig.~\ref{fig:fg6}.  The dispersion curve corresponding to $M_{\rm A} = 1.315$ divides the rest curves into two types, notably ``simple'' dispersion curves (for $M_{\rm A} < 1.315$) and much complex families of dispersion curves (for $M_{\rm A} > 1.315$).
  \label{fig:fg7}}       
\end{figure}

The case with sausage Hall-MHD waves is similar in many ways, although there are some specific features.  First, for positive relative Alfv\'enic Mach numbers the real solutions in the long-wavelength limit (small $K$s) are much more complicated.  That can be seen in Figs.~\ref{fig:fg6} and \ref{fig:fg7}.  The most striking issue is the existence of a bordering dispersion curve (with label `1.315') which divides the rest dispersion curves into two types.  The dispersion curves with $M_{\rm A} < 1.315$ begin as super-Alfv\'enic waves with decreasing phase velocities which passing through a minimum start to grow and at around $K \approx 5$ cross the neutral dispersion curve (that corresponding to a static slab).  After that they quickly increase their magnitudes reaching very large values at $K \to K_{\rm limit}$.  The second type of dispersion curves consist of two families of curves: ones at very narrow regions of small $K$s, and others starting with negligibly small negative phase velocities which later on passing through inverted ``s-shaped'' parts continuously increase their speeds reaching large values at $K \to K_{\rm limit}$.  It is worth noticing that the second type's dispersion curves possess multiple values at a fixed $K$ in the range between $0$ and $1.22$.

\begin{figure}[!hb]
\centering\includegraphics[height=.30\textheight]{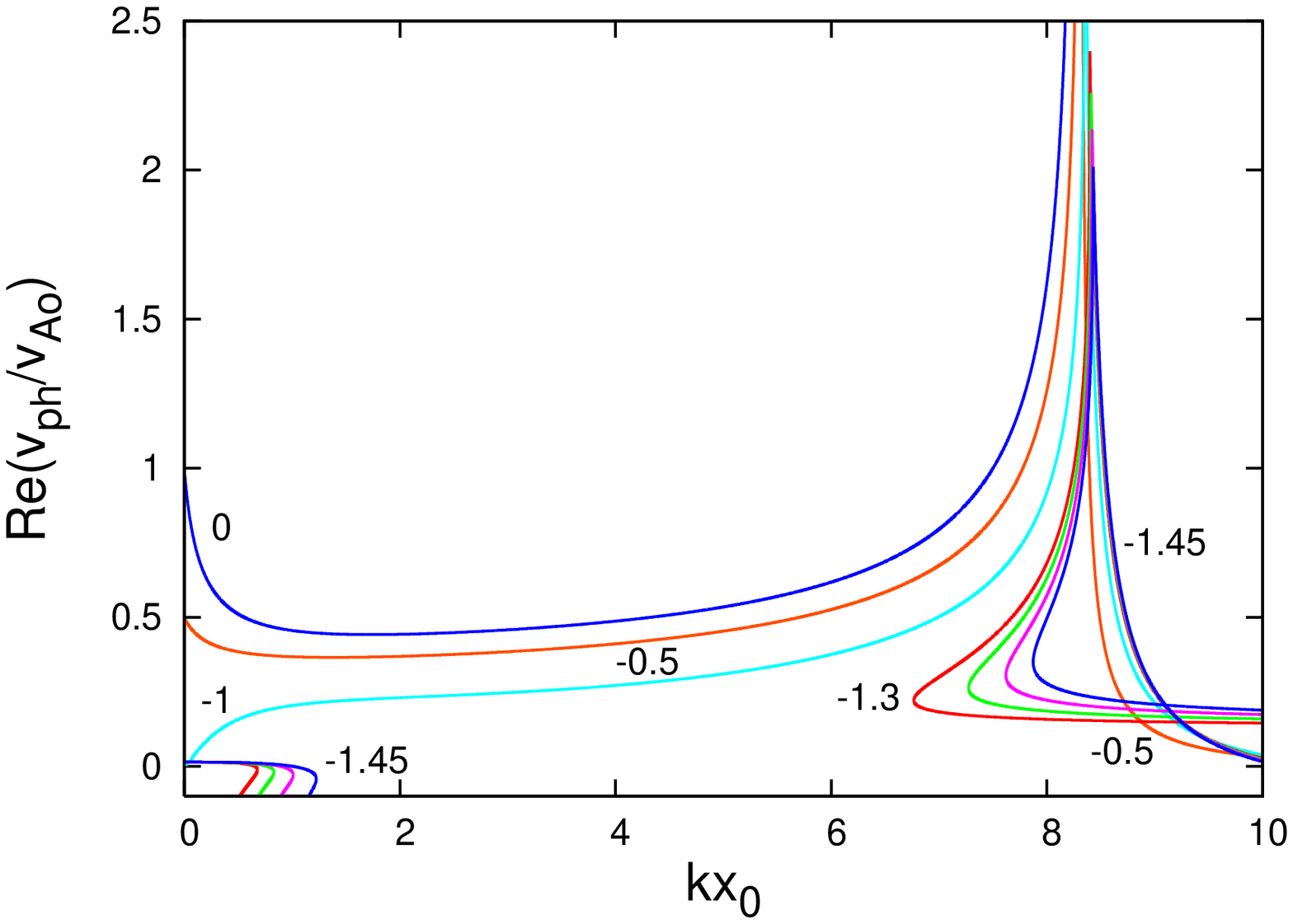}
  \caption{(Online colour) Dispersion curves of sausage Hall-MHD waves travelling along an incompressible flowing plasma layer for negative values of the relative Alfv\'enic Mach number $M_{\rm A}$.  All the curves lying on the left side of the vertical merging line/cusp represent stable waves' propagation.  The waves become unstable only on the family of curves labeled by `$-0.5$' and `$-1.45$'.  For details in curves' labeling see Fig.~\ref{fig:fg9}.
  \label{fig:fg8}}       
\end{figure}

\begin{figure}[!ht]
\centering\includegraphics[height=.30\textheight]{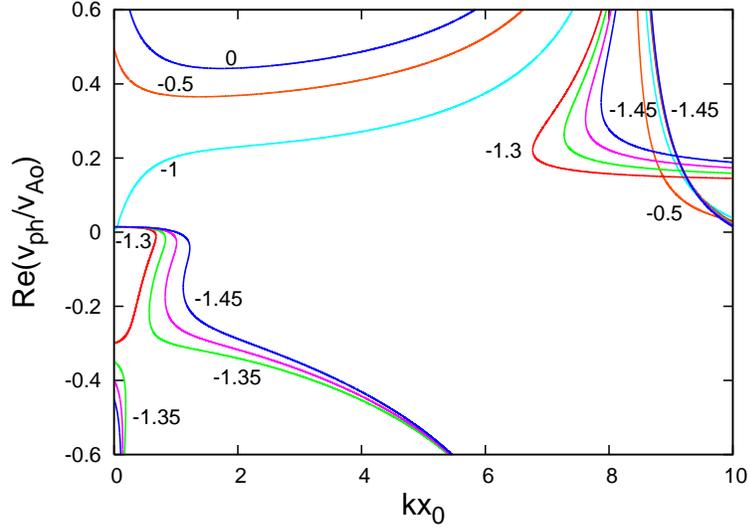}
  \caption{(Online colour) A zoom of the bottom part of Fig.~\ref{fig:fg8}.  The dispersion curves sandwiched between the curves labeled by `$-1.3$' and `$-1.45$' correspond to $M_{\rm A}$ equal to $-1.35$ and $-1.4$, respectively.  The curves situated at the left bottom corner of the dispersion diagram represent backward stable sausage surface waves.
  \label{fig:fg9}}       
\end{figure}

\begin{figure}[!hb]
\centering\includegraphics[height=.30\textheight]{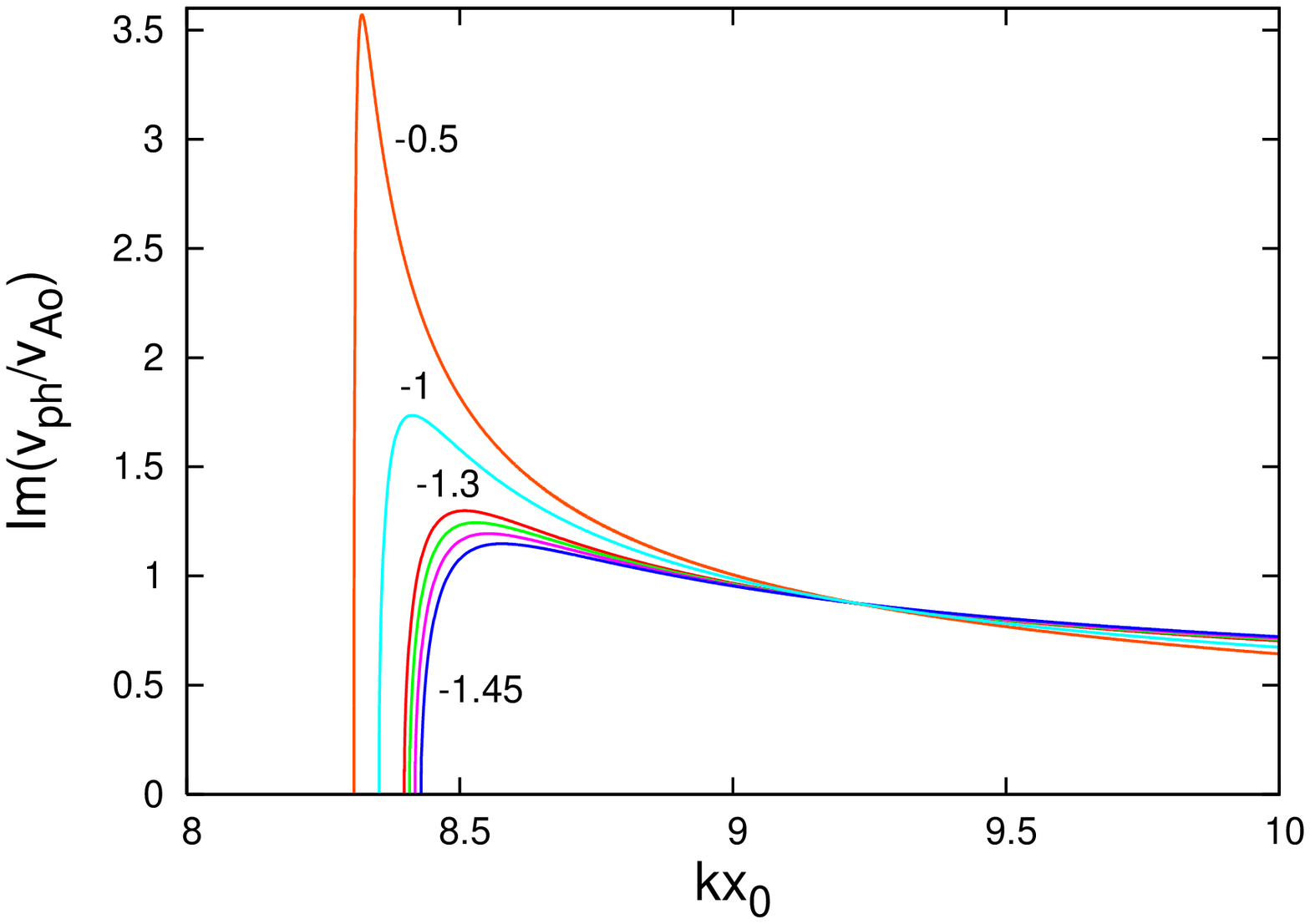}
  \caption{(Online colour) Growth rates of unstable sausage Hall-MHD waves travelling along an incompressible flowing plasma layer for negative values of the relative Alfv\'enic Mach number $M_{\rm A}$ in the short-wavelength region (beyond the $K_{\rm limit}$).
  \label{fig:fg10}}       
\end{figure}

The most intriguing question is how the dispersion curves will behave as $M_{\rm A}$ is negative?  The answer is illustrated in Figs.~\ref{fig:fg8} and \ref{fig:fg9}.  Actually there is no a big surprise -- more or less the dispersion curves are similar to those of the kink mode.  However, we should immediately notice two differences: (i) the dramatic changes in the shapes of the dispersion curves start at $M_{\rm A} = -1.3$ (vs.\ $-1.25$ for the kink mode), and (ii) parts of dispersion curves for $M_{\rm A} \leqslant -1.3$ with dimensionless wave number between $0$ and $8.2$ are negative.  All these forward and backward sausage waves are stable.  Unstable are only those waves (like for the case of kink waves) whose dispersion curves are on the right of the merging vertical line (look at Fig.~\ref{fig:fg8}).  The growth rates of such unstable sausage Hall-MHD modes are shown in Fig.~\ref{fig:fg10} and each of them starts at some critical normalized wave number.

We should recall that our theory is a linear one and since at the instability onset the waves amplitudes begin rising, for a further waves' evolution one must employ a nonlinear approach.  Nevertheless, results, obtained here, can be used as a start point for a deeper investigation of the wave propagation in flowing solar flux-tube plasmas in the framework of the Hall magnetohydrodynamic.

\begin{figure}[!ht]
\centering\includegraphics[height=.30\textheight]{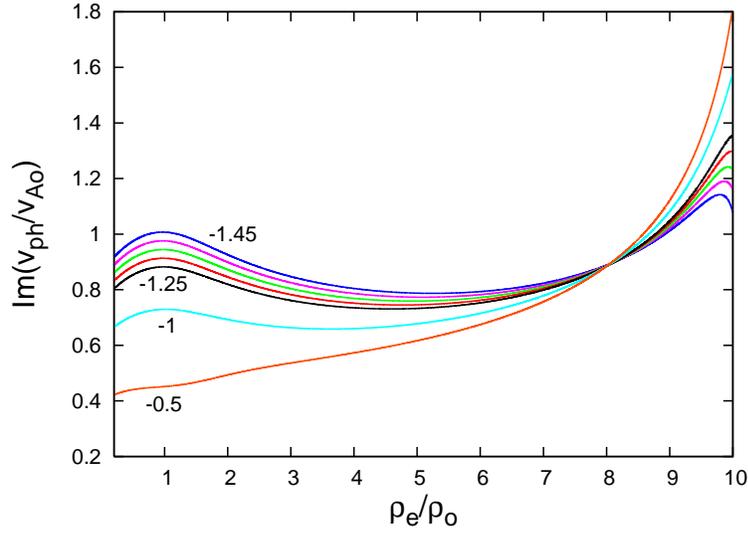}
  \caption{(Online colour) Dependence of the normalized instability growth rate on the ratio $\rho_{\rm e}/\rho_{\rm o} = \eta$ at a fixed dimensionless wave number ($kx_0 = 8.5$) for various relative Alfv\'enic Mach numbers in the range between $-1.45$ and $-0.5$  The beginning of the horizontal axis starts at $\eta = 0.2$.  The family of curves sandwiched between the curves labeled by `$-1.25$' and `$-1.45$' correspond to relative Alfv\'enic Mach numbers equal to $-1.3$, $-1.35$, and $-1.4$, respectively.
  \label{fig:fg11}}       
\end{figure}

An interesting issue which springs to mind is how the instability growth rate depends on the value of the entry parameter $\eta$, say, at a fixed dimensionless wave number.  Let us do such an examination for the kink mode and let our choice for the fixed wave number be (see Fig.~\ref{fig:fg5}) $kx_0 = 8.5$.  The wave growth rate depends not only on $K$ but also on the relative Alfv\'enic Mach number $M_{\rm A}$.  In Fig.~\ref{fig:fg11} we show a family of curves depicting the dependence of the normalized wave growth rate as a function of the plasma densities ratio $\eta = \rho_{\rm e}/\rho_{\rm o}$ for various Alfv\'enic Mach numbers between $-1.45$ and $-0.5$.  As seen, one observes two local maxima: one at $\eta = 1$ and another around $\eta \sim 10$.  The first maximum is not surprising -- it is well known that the Kelvin--Helmholtz instability is easily excited when the densities of the two adjacent flowing media are approximately the same.  The only exception here is the curve corresponding to $M_{\rm A} = -0.5$ -- a maximum of the wave growth rate for that value of the Alfv\'enic Mach number one can expect for values of $\eta$ bigger than $10$.  The second local maxima are obviously depending on the magnitude of the relative Alfv\'enic Mach number $M_{\rm A}$.

\begin{figure}[!hb]
\centering\includegraphics[height=.30\textheight]{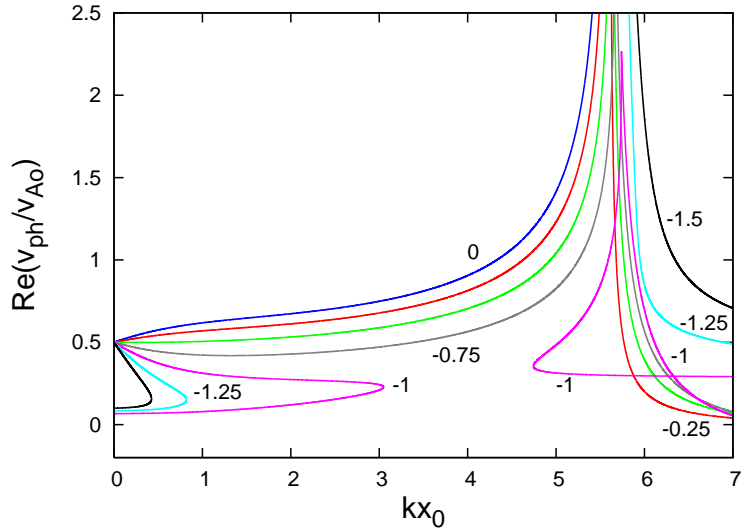}
  \caption{(Online colour) Dispersion curves of kink Hall-MHD waves travelling along an incompressible flowing plasma layer for negative values of the relative Alfv\'enic Mach number $M_{\rm A}$ and $\eta = 4$.  All the curves lying on the left side of the vertical merging line/cusp represent stable waves' propagation.  The waves become unstable only on the family of curves labeled by `$-0.25$' and `$-1$'.  The curves with labels `$-1.25$' and `$-1.5$' represent stable waves' propagation.
  \label{fig:fg12}}       
\end{figure}

\begin{figure}[!ht]
\centering\includegraphics[height=.30\textheight]{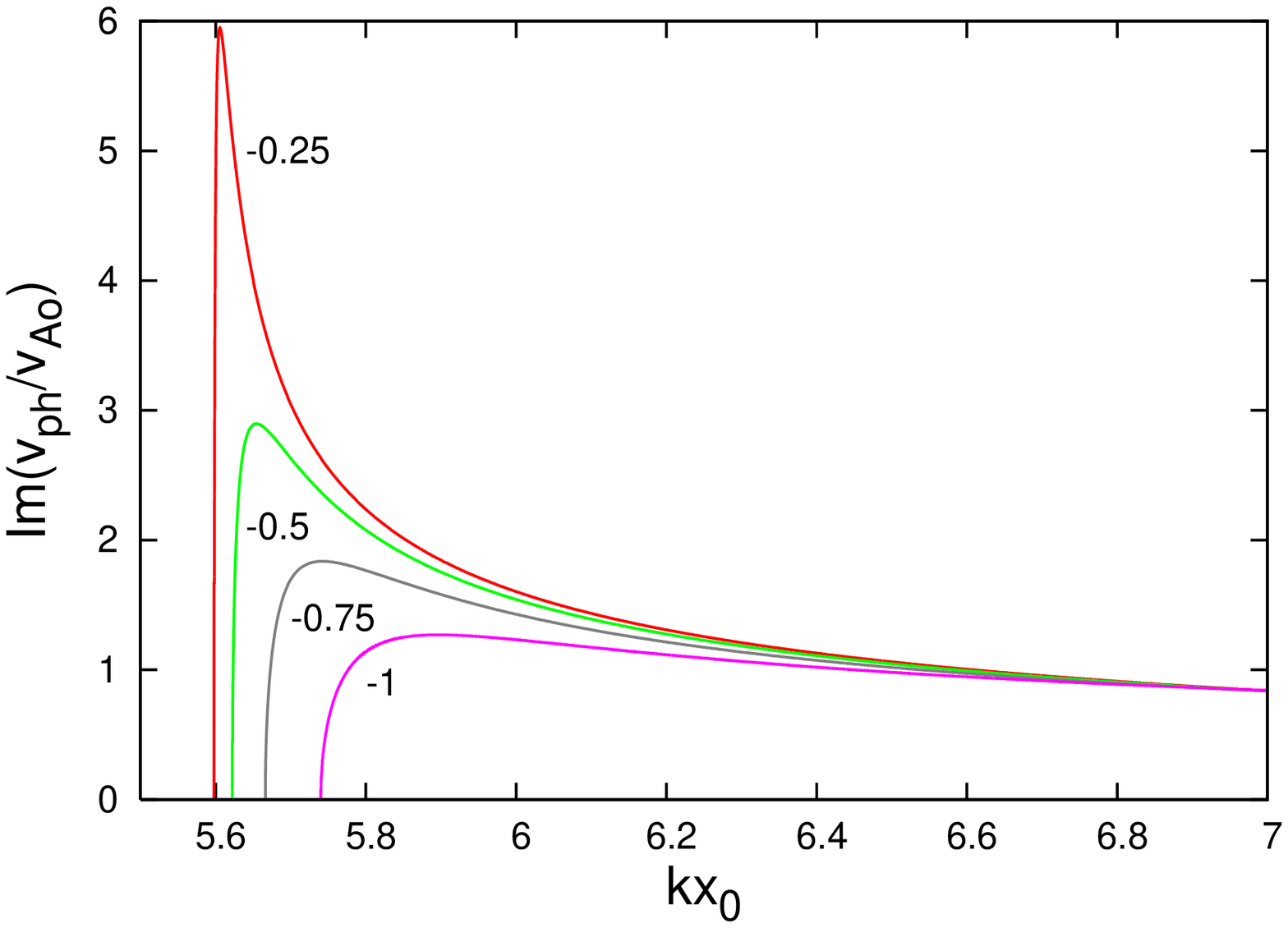}
  \caption{(Online colour) Growth rates of unstable kink Hall-MHD waves travelling along an incompressible flowing plasma layer for negative values of the relative Alfv\'enic Mach number $M_{\rm A}$ and $\eta = 4$ in the short-wavelength region (beyond the $K_{\rm limit} = 5.59$).
  \label{fig:fg13}}       
\end{figure}

Another curious question is how the propagation and stability properties of the Hall-MHD surface waves change with the value of $\eta$.  If we take $\eta = 4$, the dispersion curves of kink waves for negative values of the relative Alfv\'enic Mach number are plotted in Fig.~\ref{fig:fg12}.  As seen, the picture is similar to that shown in Fig.~\ref{fig:fg4}, however, with an distinctive feature, notably the curves corresponding to $M_{\rm A} < -1$ represent stable wave propagation even on the right side of the vertical merging line (look at curves labeled by `$-1.25$' and `$-1.5$', respectively).  This really surprising observation indicates that the Hall-MHD kink surface waves for that value of $\eta$ ($=4$) are unstable only for negative relative Alfv\'enic Mach numbers $M_{\rm A} \geqslant -1$ -- their normalized growth rates are plotted in Fig.~\ref{fig:fg13}.  This example shows us that we must be very cautious with stating some very general criteria for the Kelvin--Helmholtz instability onset with Hall-MHD waves propagating in flowing solar flux-tube plasmas.

\section{Conclusion and outlook}
\label{sec:concl}

Let us now summarize the basic results obtained in this study.  In investigating the wave propagation along a jet moving with respect to the environment with a constant speed $U$ we had to take into account the influence of two factors: (i) the Hall term in the generalized Ohm's law, and (ii) the flow itself.  The combining effect of these two factors can be expressed as follows:
\begin{itemize}
  \item The Hall term generally limits the range of propagation of the wave modes not only for static tubes/layers but also for jets with positive Alfv\'enic Mach numbers.  The limiting normalized wave number, $K_{\rm limit}$, is specified by two plasma parameters: the densities ratio of the two plasma media (outside and inside the jet), $\eta$, and the Hall parameter, $\varepsilon$ [look at Eq.~(\ref{eq:klimit})].  If this is the exact value for the waves propagating on a static flux tube, it is approximately the same for the waves travelling along a jet.  One should emphasize that all dispersion curves, at relatively large $K$s, lie on the left to the dispersion curve corresponding to $M_{\rm A} = 0$ (see Figs.~\ref{fig:fg1} and \ref{fig:fg6}).  However, when $M_{\rm A}$ becomes negative the real part of the phase velocity of the eigenmodes (kink and sausage waves) is forced to go beyond that limiting wave number in a region where the wave becomes unstable (or if you prefer, overstable).  The instability which occurs should be of Kelvin--Helmholtz type.  It is rather surprising that the instability onset starts with relatively large growth rates gradually decreasing with increasing the modulus of the Alfv\'enic Mach number (look at Figs.~\ref{fig:fg5} and \ref{fig:fg10}).  We note that such growth rates in the extended range of the waves' propagation was obtained for the same plasma-jet configuration for a different value of the parameter $\eta$ ($= 0.64$) \cite{ivan07}.  It is worth mentioning that for negative Alfv\'enic Mach numbers we actually have (for each mode) two dispersion curves merging at $K_{\rm limit}$, but unstable for negative $M_{\rm A}$ are the dispersion curves situated on the right to the cusp (see Fig.~\ref{fig:fg3} and \ref{fig:fg8}).  Our conclusion that the Kelvin--Helmholtz instability onset is only possible for negative relative Alfv\'enic Mach numbers is in agreement with a similar inference of Andries and Goossens \cite{andries01}.

  \item It seems that the critical relative Alfv\'enic Mach number at which the Kelvin--Helmholtz instability starts, given by Eq.~(\ref{eq:eta}), is unfortunately not applicable, as a rough estimation, in the Hall magnetohydrodynamics.  Notwithstanding, it is still useful because yields that value of the negative Alfv\'enic Mach number which is associated with a dramatic change in the shape of the waves' dispersion curves.

  \item A safely general conclusion is that the Hall current keeps stable the surface modes travelling in flowing solar plasmas within the dimensionless wave number range between $0$ and $K_{\rm limit}$ for each relative Alfv\'enic Mach number.  An instability of the Kelvin--Helmholtz type is possible only at negative Alfv\'enic Mach numbers in a wave number range lying beyond the $K_{\rm limit}$ and it (the instability) starts at some critical normalized wave number depending on the magnitude of $M_{\rm A}$. The maximum instability growth rate is largest for small in magnitude Alfv\'enic Mach numbers gradually decreasing with the increase of $\left|M_{\rm A}\right|$.  The instability can turn off as $\left|M_{\rm A}\right|$ reaches some value depending on the magnitude of the parameter $\eta$ (look at Figs.~\ref{fig:fg12} and \ref{fig:fg13}).
\end{itemize}

This study can be extended in two directions.  The first one is to consider finite-valued sound speeds.  In that case, however, the waves' dispersion equations become rather complicated and they can be solved (looking for complex roots) only numerically which is a highly difficult task.  Any way, one can expect that the plasma compressibility will not substantially change the Kelvin--Helmholtz  instability pictures.  Moreover, one can state that the sausage and kink waves represented in Figs.~5 and 7 in Ref.~\cite{rossi04} are definitely stable because with $\eta \approx 0.6$ and $\varepsilon = 0.4$ the value of $K_{\rm limit}$, beyond  which one can expect the instability onset, is $3.16$, while the waves' propagation in that paper was examined till $kx_0 = 2.5$ only.
The second direction is to study a more realistic geometry, for instance, cylindrical one, and conduct the investigations with appropriate observable plasma and magnetic field parameters.  This is in progress and will be reported elsewhere.

We do believe that the present results might be useful in studying wave turbulence in the solar wind as well as in solving other problems associated with wave propagation in structured/spatially bounded magnetized plasmas.

\section*{Acknowledgments}

The author would like to thank Michael Ruderman for useful discussions and advices, as well as the referee for his valuable critical remarks and helpful suggestions.

\clearpage
\large{\centerline{\bfseries{\textsf{Figures and Figure Captions}}}}
%
%
%

\end{document}